# Synthesizing Skyrmion Bound Pairs in Fe-Gd Thin Films

J. C. T Lee,[1, 2, 4, a)] J. J. Chess,[1, a)] S. A. Montoya,[3, a)] X. Shi,[1] N. Tamura,[2] S. K. Mishra,[2] P. Fischer,[4, 5] B. J. McMorran,[1] S. K. Sinha,[6] E. E. Fullerton,[3] S. D. Kevan,[1, 2, 4] and S. Roy[2, b)]

[1]*Department of Physics, University of Oregon, Eugene, Oregon 97401*

[2]*Advanced Light Source, Lawrence Berkeley National Laboratory, Berkeley, California 94720 USA*

[3]*Center for Memory and Recording Research, University of California, San Diego, California 92093, USA*

[4]*Materials Sciences Division, Lawrence Berkeley National Laboratory, Berkeley, California 94720, USA*

[5]*Physics Department, University of California, Santa Cruz, California 94056 USA*

[6]*Department of Physics, University of California, San Diego, California 92093, USA*

We show that properly engineered amorphous Fe-Gd alloy thin films with perpendicular magnetic anisotropy (PMA) exhibit bound pairs of like-polarity, opposite helicity skyrmions at room temperature. Magnetic mirror symmetry planes present in the stripe phase, instead of chiral exchange, determine the internal skyrmion structure and the net achirality of the skyrmion phase. Our study shows that stripe domain engineering in amorphous alloy thin films may enable the creation of skyrmion phases with technologically desirable properties.

---

[a)] J. C. T Lee, J. J. Chess, and S. A. Montoya contributed equally to this work.
[b)] Author to whom correspondence should be addressed. Electronic mail: sroy@lbl.gov

Fifty years since Skyrme described baryons as topological defects of continuous fields[1], excited "skyrmion" states have been found in condensed matter systems such as liquid crystals[2], quantum Hall systems[3], ferroelectrics[4], and magnetic materials[5]. Similar excitations were theoretically explored in two-dimensional magnetic systems, which evade ferromagnetic transitions by developing spatially inhomogeneous topological spin defects.[5, 6, 7, 8] The topology is described by a quantized and conserved skyrmion number. The spin texture topology also "protects" skyrmions from scattering by structural defects, allowing them to be moved with $\sim 10^5$ times lower current density than a conventional magnetic domain[9]. These features make magnetic skyrmions appealing for low power memory and information processing applications based on spin torque transfer and the topological spin Hall effect.[10]

Topological magnetic skyrmions have now been observed in several bulk and thin film systems.[8, 11, 12, 13, 14, 15, 16] Magnetic skyrmion phases based on chiral exchange result from a competition between chiral exchange, e.g., Dzyaloshinskii-Moriya interactions (DMI) and symmetric exchange. DMI is a relatively weak broken symmetry in the magnetic Hamiltonian and are generally observed well below room temperature.

Locally, chiral textures can also arise without chiral interactions. For example, bent-core liquid crystals exhibit phases with domains of opposite chirality despite the achirality of the molecules.[17] In magnetic systems, skyrmions driven by achiral dipolar interactions are predicted.[18] Indeed, topologically trivial bubble domains that commonly appear in systems with perpendicular magnetic anisotropy (PMA) above room temperature can be described by the same effective field theories as skyrmions. An important step in skyrmion physics will be to develop ways to produce and control specific topological phases in achiral materials, like the scheme illustrated in Refs. 19, 20, 21, and 22.

Using resonant soft x-ray scattering (RSXS) and Lorentz transmission electron microscopy (LTEM), we have discovered and characterized a skyrmion phase that exists up to 300 K in amorphous Fe-Gd PMA thin films. This phase is composed of bound pairs of unit winding number skyrmions with opposite helicity and aligned polarity. To produce the bound pair phase, we applied a magnetic field with a small in-plane component while rapidly decreasing its out of plane component to zero. This creates a magnetic stripe texture that is crucial to producing the skyrmion bound pair phase. While the skyrmion bound pair phase is similar to that seen[12] in centrosymmetric $La_{1+2x}Sr_{2-2x}Mn_2O_7$, our report concerns their creation in amorphous thin films at room temperature. The skyrmion bound pair phase is very sensitive to film structure, especially thickness, as well as alloy composition and magnetization protocol. This means that Fe-Gd films are an easily tunable, room temperature thin film skyrmion system.

The samples studied were nominally [Gd (0.4 nm)/Fe (0.34 nm)]×80 multilayers deposited using DC magnetron sputtering with 20 nm Ta seed and capping layers. The samples were deposited on 50-nm or 200-nm thick $Si_3N_4$ membranes to allow for LTEM and transmission RSXS experiments, respectively. Non-resonant 12 keV x-ray diffraction from the films, conducted at Beamline 12.3.2, Advanced Light Source (ALS), Lawrence Berkeley National Laboratory, indicated that the Gd and Fe layers are strongly intermixed, forming an amorphous structure rather than a multilayer. The lack of a crystal lattice with a consistent, repeating local inversion symmetry breaking to point DMI vectors in one direction; the lack of Fe/Gd interfaces; and the weak effect of Rashba spin-orbit effects in films as thick as 60 nm indicate that there is no chiral symmetry breaking interaction in the Fe-Gd films.

RSXS experiments with Fe $L_3$ (≈707 eV) and Gd $M_5$ (≈1198 eV) x-rays were performed at Beamline 12.0.2.2 at the ALS to probe the film's out-of-plane magnetic structure in momentum space. A flow cryostat controlled the sample temperature, which we changed in zero-field conditions. RSXS data were obtained at fixed temperatures during magnetic field cycles.

A minimum magnetic field of ~ 1 mT was applied parallel to the film's surface at all times, while we changed only the out-of-plane component of the applied field. The out-of-plane magnetic field cycles proceeded in steps, from zero-field to $B_{max}$ = -500 mT (directed antiparallel to the beam); then from $-B_{max}$ to $+B_{max}$ (directed parallel to the beam); then finally from $+B_{max}$ to zero-field. With the exception of the $+B_{max}$ to zero-field portion of the loop, the field time rate of change from one step to the next was $\Delta B/\Delta t \leq$ 13 mT/s. During the $+B_{max}$ to zero-field portion of the loop, $\Delta B/\Delta t \approx$ 380 mT/s. At lower rates of change, the linear stripes disorder into labyrinthine stripes with repeated field cycling.

LTEM was conducted at the University of Oregon using an FEI Titan microscope to map the real space structure of the in-plane magnetic structure.

The Gd and Fe moments are ferrimagnetically aligned and, for the nominal layer thicknesses and temperatures studied here, the Gd provides the dominant moment. RSXS measurements were performed starting from the magnetic stripe phase observed at 300 K and remanence. The remanent stripe period is 125 ± 3 nm. Odd-order diffraction peaks dominated in the remanent state, indicating that the reversal domains have nearly equal widths. Applied out-of-plane magnetic fields ($H_{OOP}$) broke this equality, causing even order diffraction peaks to appear. (See Figure 1(a).) At $\mu_0 H_{OOP}$ = 154.0 ± 2.0 mT, the intensity of all orders started to decrease and eventually new peaks appeared in the form of an approximately hexagonal pattern (see Fig. 1(b)). At higher field, the nominally hexagonal scattering pattern became more distinct with the appearance of higher order peaks (see Figs. 1(c) and 1(d)). At still higher fields, the film became uniformly magnetized and all diffraction peaks disappeared. This behavior is qualitatively similar to neutron scattering results from other skyrmion materials.[8] At lower temperatures, the stripe peaks broadened and the stripe-hexagonal transition became less distinct. At temperatures below 217 K, the transition failed to occur and only disordered stripes were observed. (Scattering patterns at lower temperature are shown in the Supplemental Material.[23])

Figure 1(e) shows a temperature-magnetic field phase diagram summarizing our scattering data. The hexagonal lattice phase, stripe phase, and a transition region between these two appear as blue, red, and grey areas, respectively. The vanishing of the hexagonal lattice order parameter at low temperature does not follow the accepted mean field theory description of magnetic bubble systems, which predicts persistence of bubbles to T=0.

Although the diffraction patterns appear to reflect a hexagonal lattice, the pronounced intensity variations of nominally symmetry equivalent peaks (e.g., peaks 10 and 01; peaks 12 and 21, in Fig. 1(d)) reveals that the out-of-plane magnetization has a lower two-fold symmetry. Since skyrmions, regardless of winding number (Z), have azimuthally symmetric out-of-plane magnetization,[3] we exclude a hexagonal lattice of individual skyrmions of any Z. A plausible two-fold symmetric basis is a skyrmion bound pair. Bound skyrmion pairs have also been shown, under certain conditions, to be the lowest energy state in quantum Hall systems.[3] Similar bound pairs have been observed in ferromagnetic manganites and referred to as biskyrmions.[12] Confusingly, single skyrmions with Z=2, which have spin textures similar to bound pairs, are also called biskyrmions. We will return to the distinction between skyrmion bound pairs and biskyrmions.

Stripe-skyrmion coexistence cannot explain the lowered symmetry we observed, since it persists in fields high enough to eliminate the remaining stripes. Moreover, the diffraction pattern from a stripe-skyrmion structure would simply look like a sum of separate stripe and skyrmion lattice diffraction patterns. Instead, we observe a systematic breaking of six-fold symmetry in the RSXS pattern consistent with a lattice of two-fold symmetric objects. When the stripes break up to form skyrmion bound pairs, a hexagonal space group is not allowed. However, dipole interactions between skyrmion bound pairs will still favor efficient packing, and a distorted hexagonal lattice with a single two-fold symmetry axis is formed. Since the most intense lowest order peaks of the skyrmion phase are aligned with the original stripe peaks, the axis joining the skyrmions must be approximately perpendicular to the original stripe domains.

Further insight about the skyrmion bound pair comes from LTEM measurements that map the real space in-plane magnetization. A series of images was obtained at room temperature starting from the remanence condition and to saturation. At the field where the skyrmion lattice was observed with RSXS, the LTEM image in Fig. 2(a) shows the in-plane magnetic induction, which we reconstructed using the transport of intensity equation (TIE), of the roughly hexagonal lattice of bound pairs. The red-colored long axes of the bound pairs are oriented in the same direction as the stripes throughout. As denoted by the colors of the bound pairs, the in-plane fields of the skyrmion bound pairs swirl in opposite directions, reflecting their opposite helicities. This is made explicitly clear in Fig. 2(b), which shows a vector map of the magnetic induction.

In Fig. 3, we schematically illustrate the magnetic stripe-to-skyrmion transition as a function of applied field, based on a careful analysis of our resonant x-ray diffraction and LTEM results. We find that the common picture of stripe domains in a PMA system breaks down in this system. Specifically, there are (relatively) wide regions of canted spins, dark green and pink in Fig. 3(a), separated by narrow regions where spins rotate rapidly in space, shaded bright green. Some of the spins in the latter regions align perpendicular to the film and ultimately form the skyrmion cores. The in-plane components of the dark green and pink spins are aligned parallel and antiparallel, respectively, to the in-plane field direction applied throughout the reversal process. We note that the spin direction reverses its sense of rotation twice per period, and therefore has no net chirality.[12] Without an in-plane field, the chirality of each domain wall will be random, making it difficult for the stripes to form structures such as skyrmion bound pairs, the winding numbers of which add to zero.

Increasing the out-of-plane applied field expands the dark up-spin regions, rotates the bright green spins toward the field direction, and the pink spins to rotate increasingly toward the plane of the film (Fig. 3(b)). These in-plane spins are captured between two narrow down-spin regions, shaded white. When their width becomes comparable to the domain wall width, the stripes become unstable and break into shorter segments. This occurs over a narrow region of applied magnetic field, which explains why the skyrmion bound pair phase is supported over a limited range of applied fields. At higher applied fields, the stripes break up and the dipolar fields of the pink in-plane segments induce the formation of swirling spin textures, indicated by the multi-hued arcs in Fig. 3(c). The pink segments form a distorted hexagonal lattice to minimize dipolar energy. The white spins, originally adjacent to the pink domains in Fig. 3(b), are surrounded by the swirling spin texture and become the cores of the nascent skyrmions. This spin texture is that of two skyrmions with opposite helicities (thus, no net chirality) but same central spin polarity, as observed in our results.

Finally, we note that in Fig. 2 the regions of in-plane spin in the individual skyrmion bound pairs align nearly parallel to one another in the skyrmion lattice. This orientation can be traced back to the in-plane component of the applied field, which leads to two mirror planes perpendicular to the individual stripes and paired chirality reversals. We conclude that the formation of skyrmion bound pairs is driven by the broken symmetry of the in-plane field, not from a chiral DM interaction, resulting in the achiral skyrmion phase we observe.

We point out that the skyrmion bound pairs found in Fe-Gd films and in manganites[12] differ from single $Z=2$ skyrmions. Despite their spin textures having similar in-plane features and the same overall skyrmion number ($Z\times$polarity), bound pairs and $Z=2$ skyrmions are fundamentally different. First, their spin textures are dissimilar since the out-of-plane component of bound pair magnetization is not azimuthally symmetric like that of a $Z=2$ skyrmion.[23] Further, a bound pair is a composite of topological objects, whereas a $Z=2$ skyrmion is a unitary topological object. In principle, a bound pair can be smoothly separated into two separate topological units by an applied magnetic field of appropriate strength, whereas a $Z=2$ skyrmion cannot.

Skyrmion bound pairs have been observed in numerical simulations of quantum Hall systems.[3] Aside from the effects of the demagnetizing field on the effective Zeeman term in Fe-Gd thin films, quantum Hall systems are analogs of our films. Drawing on the analysis of quantum Hall skyrmions, we conclude that relatively small exchange and anisotropy constants (resulting in low domain wall energies) are required to obtain skyrmion phases. These parameters can be tuned by varying elemental composition and material thickness.

The transition from stripe domains to bubble states in PMA materials has been extensively studied.[24,25] The new development presented here is the topological aspect. While many bubble materials show bubbles of zero winding number, they can also develop skyrmions with nontrivial winding numbers. We conclude that skyrmion bound pairs in Fe-Gd thin films can be synthesized by controlling the composition and by manipulating the magnetic field history.

Work at the ALS, LBNL was supported by the Director, Office of Science, Office of Basic Energy Sciences, of the US Department of Energy (Contract No. DE-AC02-05CH11231). Work at the University of Oregon was partially supported by the U.S. Department of Energy, Office of Basic Energy Sciences, Division of Materials Science and Engineering under Grant No. DE- FG02-11ER46831. The research at UCSD was supported by the research programs of the US Department of Energy (DOE), Office of Basic Energy Sciences (Award No. DE-SC0003678). S.K. and P.F. were supported by the Director, Office of Science, Office of Basic Energy Sciences, Materials Sciences and Engineering Division, of the U.S. Department of Energy under Contract No. DE-AC02-05-CH11231 within the Non-Equilibrium Magnetic Materials Program at LBNL. B.J.M. and J.J.C. gratefully acknowledge the use of the CAMCOR High Resolution and Analytical Facility at University of Oregon, which is supported by the W.M Keck Foundation, the M.J. Murdock Charitable Trust, ONAMI, the Air Force Research Laboratory (Agreement No. FA8650-05-1-5041), NSF (Award Nos. 0923577, 0421086) and the University of Oregon.

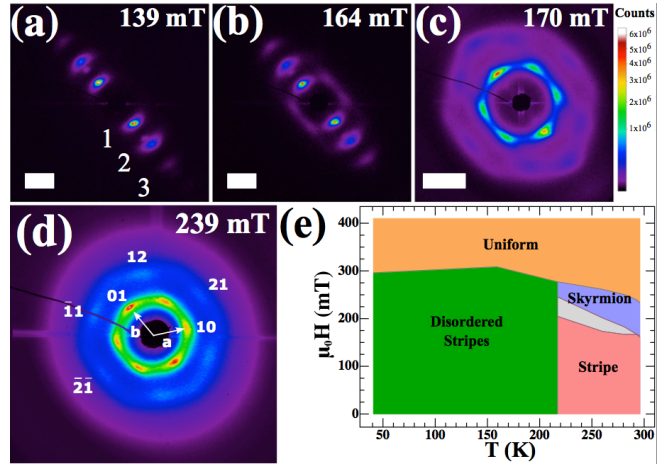

**Figure 1** Representative Gd $M_5$ resonant scattering patterns, square root scale. Scale bars represent a transfer momentum of 0.041 nm$^{-1}$. (a) Scattering from stripe domains with the first, second, and third order stripe peaks visible. Parts (b)-(d) illustrate evolution of skyrmion bound pair lattice with magnetic field. In part (d), reciprocal basis vectors **a** and **b** are shown and the diffraction peaks are labeled by Miller indices. Note that peaks 10 and 01 are not equally intense; and that peaks 12 and $\bar{1}1$ are more intense than peaks 21 or $\bar{2}\bar{1}$. (e) Temperature-magnetic field phase map. Grey region represents a broadened stripe-skyrmion transition. The abrupt phase boundary at 217 K is due to coarser temperature steps in that region. Scattering patterns from "Disordered Stripes" and the transition region are illustrated in the Supplemental Material.[23]

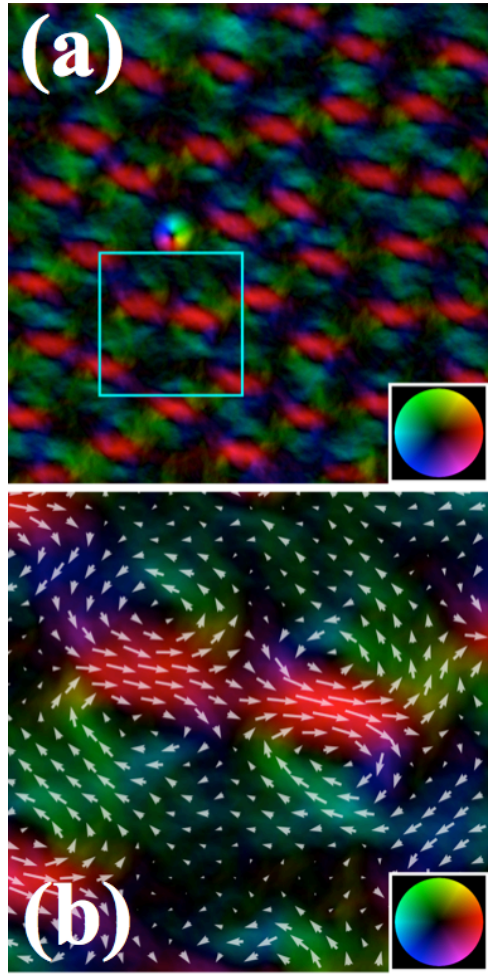

**Figure 2** In-plane component of magnetic induction (**B**) of skyrmion bound pairs obtained using the TIE method from LTEM images. Obtained at room temperature and $\mu_0 H = 207$ mT. The color wheels relate the color to the in-plane orientation of **B**. (a) Bound pairs are the majority of the objects in the image, with red lines running through their centers. An isolated (circular) unit winding number skyrmion is seen. (b) A vector map of the $0.5 \times 0.5$ $\mu m^2$ boxed region in part (a). The **B**-fields of the top and bottom halves of the bound pair (above and below the red center) have opposite circulations.

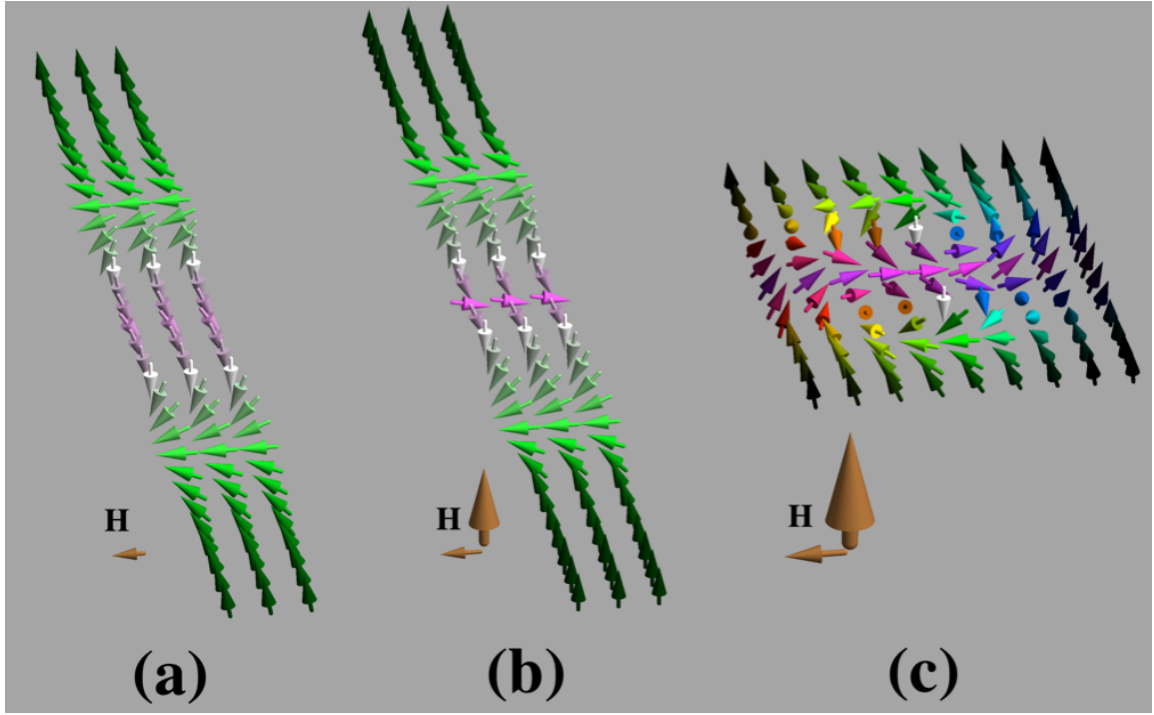

**Figure 3** Schematic evolution from canted stripes to the skyrmion bound pairs. Color and brightness denote direction and magnitude of spin in-plane component; green and pink have opposite in-plane directions. Brown arrows denote the applied magnetic field, the vertical arrow being the out-of-plane component and the horizontal arrow being the in-plane component. (a) In low fields, up and down stripes are nearly of equal width. The rotational sense of the magnetization switches at the center of these regions. (b) Applied field causes dark green regions to widen, pink regions to narrow, and tilts spins upward. (c) At a critical field, the stripes pinch off into short patches. The dipole fields of these patches cause green spins in parts (a) and (b) to form swirling (multi-hued) spin textures and out-of-plane (white) spins to form the cores of nascent skyrmions.

# Supplemental Information for "Synthesizing Skyrmion Bound Pairs in Fe-Gd Thin Films"

J. C. T Lee,[1,2,4,a)] J. J. Chess,[1,a)] S. A. Montoya,[3,a)] X. Shi,[1] N. Tamura,[2] S. K. Mishra,[2] P. Fischer,[4,5] B. J. McMorran,[1] S. K. Sinha,[6] E. E. Fullerton,[3] S. D. Kevan,[1,2,4] and S. Roy[2,b)]

[1]*Department of Physics, University of Oregon, Eugene, Oregon 97401*

[2]*Advanced Light Source, Lawrence Berkeley National Laboratory, Berkeley, California 94720 USA*

[3]*Center for Memory and Recording Research, University of California, San Diego, California 92093, USA*

[4]*Materials Sciences Division, Lawrence Berkeley National Laboratory, Berkeley, California 94720, USA*

[5]*Physics Department, University of California, Santa Cruz, California 94056 USA*

[6]*Department of Physics, University of California, San Diego, California 92093, USA*

This document provides Supplemental Information about the temperature dependence of resonant soft x-ray scattering observed from Fe-Gd thin films. The scattering patterns from the disordered stripe state and stripe-skyrmion transition are shown. In addition, the magnetization vector field of skyrmions of topological charges of 1 and 2 are illustrated to demonstrate the difference between skyrmion bound pairs and topological charge 2 skyrmions.

[a)] J. C. T Lee, J. J. Chess, and S. A. Montoya contributed equally to this work.
[b)] Author to whom correspondence should be addressed. Electronic mail: sroy@lbl.gov

In our letter, we describe our resonant soft x-ray scattering from the room temperature stripe and skyrmion phases in amorphous Fe-Gd thin films with perpendicular magnetic anisotropy. Here, we provide supplemental information to illustrate the characteristic scattering of the disordered stripe phase and the stripe-skyrmion transition region. In addition, we illustrate the magnetization vector fields of skyrmions with winding numbers (Z) of 1 and 2 to clarify the distinction between the skyrmion bound pairs we observe and Z=2 skyrmions.

*Temperature Dependence of Resonant X-ray Scattering*

We briefly summarize our room temperature results. The skyrmion phase is composed of bound pairs of unit winding number skyrmions with opposite helicity and like polarity. This bound pair phase arises from a well-aligned magnetic stripe texture, which we obtained by applying a magnetic field with a small in-plane component while rapidly decreasing its out of plane component to zero. This creates a magnetic stripe texture that is crucial to producing the skyrmion bound pair phase. At lower temperatures, however, the skyrmion phase occurs over a narrower range of applied magnetic fields, and the transition region between the stripe and skyrmion phases increases. At temperatures below 217 K, the transition fails to occur and only disordered stripes were observed.

Figure S1 shows in greater detail the characteristic scattering patterns found throughout the temperature-magnetic field phase diagram, including the disordered stripe and stripe-skyrmion transition regimes. Figs. S1(a)-S1(c) are room temperature data similar to that shown in the letter. The second (Fig. S1(d)-S1(f)) and third (Fig. S1(g)-S1(i)) rows illustrate the magnetic field dependence of the film at, respectively, 217 K and 159 K, temperatures at which we see disordered stripes. The phase space locations of these images are marked on the phase diagram in Fig. S1(j).

Unlike the higher temperature stripes, which form well-ordered grating-like magnetic stripe arrays, disordered stripes are broadened along a circle of constant momentum transfer ($q$). This broadening indicates that the stripes are not as straight as before but, instead, form a sinusoidally distorted[1] stripe array. This undulation becomes stronger and the scattering peaks rotate along a circle of constant-$q$ as the magnetic field increases, as shown by the progression shown in Figs. S1(d)-S1(f) and Fig. S1(g)-S1(i). At higher magnetic fields, the stripes break up and a skyrmion lattice partially forms at T=217 K, as shown in Fig. S1(f), but not at lower temperatures, as can be seen in Fig. S1(i). Interestingly, the orientation of the skyrmion lattice in Fig. S1(f) is still determined by the stripe orientation, indicating that the skyrmion bound pairs are still aligned to the local stripe direction.

We have shown that at lower-than-room temperatures, the stripe state becomes increasingly unable to form well-ordered grating-like arrays, such as those found at room temperature. Below T = 217 K, skyrmion order parameter vanishes, which does not follow the accepted mean field theory description of magnetic bubble systems, which predicts persistence of bubbles to T = 0. Mühlbauer et al.[2] found that thermally driven magnetic fluctuations stabilize skyrmion spin textures in MnSi, which is consistent with the negative slope of the stripe—hexagonal phase line observed in Fig. S1(j). This temperature dependence is probably also influenced by the rapid variation of the Gd moments with temperature. Indeed, films with a few percent lower Fe fraction form the hexagonal phase at T ~ 200 K. (Results not reported here.)

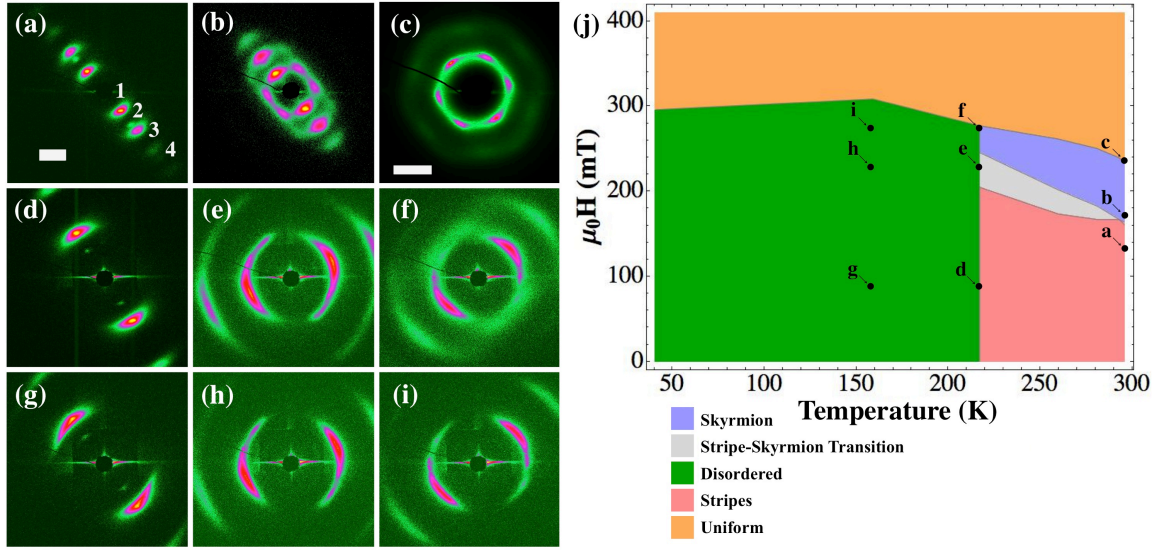

**Figure S1** Representative Gd $M_5$ resonant x-ray scattering patterns. Log scale. Parts (a)-(c) obtained at room temperature. (a) Scattering from stripe domains at $\mu_0H$ = 132 mT. The first, second, third, and fourth order stripe peaks are labeled. (The green spots near the second order peaks are due to another x-ray beam harmonic.) Parts (b) and (c) illustrate the stripe-skyrmion bound pair transition. Part (b) obtained at $\mu_0H$ = 170 mT and part (c) at 189 mT. Parts (d)-(i) illustrate magnetic domain temperature dependence: (d)-(f) T = 217 K; (g)-(i) T = 159 K. The magnetic field varies across column, left to right: $\mu_0H$ = 82 mT, 227 mT, and 271 mT. (j) Temperature—magnetic field phase map. Phase regions are labeled according to their characteristic scattering. Scale bar in part (a), which denotes a momentum transfer of 0.466 nm$^{-1}$, also applies to all scattering patterns except part (c), for which a special scale bar is shown.

*Distinction Between Skyrmion Bound Pairs and Z=2 Skyrmions*

To clarify the distinction between skyrmion bound pairs and Z=2 skyrmions, we illustrate the spin textures of skyrmions with topological charges of 1 and 2.

We base our skyrmion spin texture plots in Figure S2 on Equation 2 of Lilliehöök et al,[3] which are analytical functions describing skyrmion spin texture. The analytical functions we used:

$$n_x = -\sqrt{1-f^2(r)}\,\text{Sin}(Z\theta),\ n_y = \sqrt{1-f^2(r)}\,\text{Cos}(Z\theta),\ n_z = f(r) \quad \text{(Eq.S1)}$$

$$n_x = \sqrt{1-f^2(r)}\,\text{Cos}(Z\theta),\ n_y = \sqrt{1-f^2(r)}\,\text{Sin}(Z\theta),\ n_z = f(r) \quad \text{(Eq.S2)}$$

In Eqs. S1 and S2, Z is the integer topological charge; r and θ are respectively the radial and azimuthal coordinates measured from the skyrmion center; and $f(r) = \left[(r/\lambda)^{2Z} - 4\right]/\left[(r/\lambda)^{2Z} + 4\right]$, where λ is a characteristic skyrmion length scale. Note that $f(r)$ obeys skyrmion boundary conditions: $f(0) = -1$ and $f(\infty) = 1$.

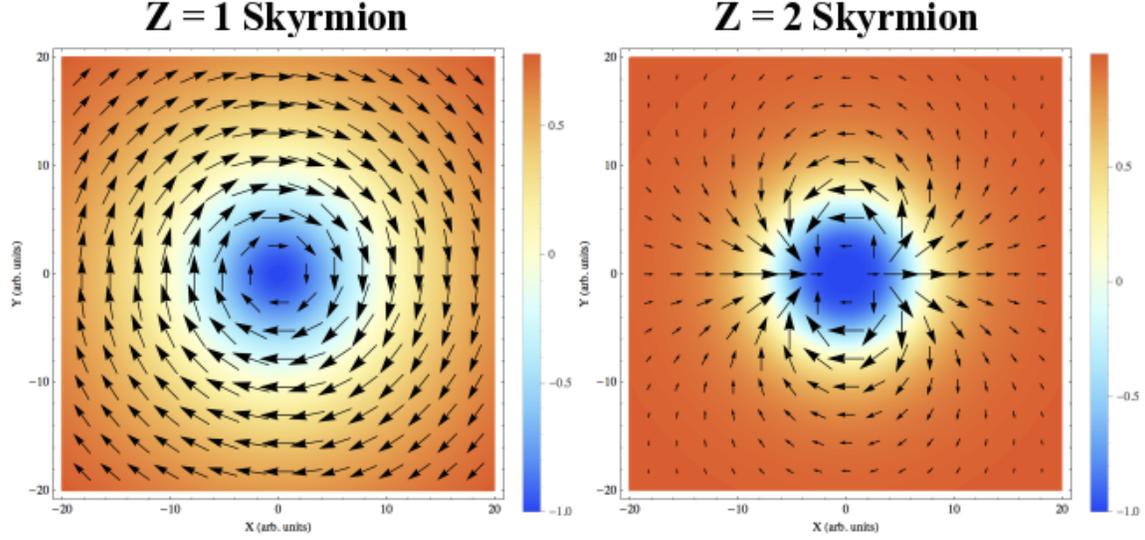

**Figure S2** Spin texture plots of skyrmions with topological charges of one (left) and two (right), based on Eqs. S1 and S2, respectively. The out of plane magnetizations of the skyrmions are azimuthally symmetric, regardless of topological charge. The color plots represent the magnitude and orientation of the out-of-plane components of the skyrmion magnetization fields. In-plane components of the magnetization field, represented by black arrows, are overlaid on the color plots. The characteristic length scales $\lambda$ of both skyrmions are the same.

Fig. S2 illustrates that skyrmions have azimuthally symmetric out of plane magnetizations, irrespective of Z. The scattering geometry of our resonant soft x-ray diffraction measurements let us distinguish between skyrmions with azimuthally symmetric out of plane magnetizations and objects that do not.

If the skyrmions in the triangular lattice of the skyrmion phase were Z=2 skyrmions rather than bound pairs of Z=1 skyrmions, no azimuthal asymmetry would appear in the magnetic diffraction pattern. Consequently, in Figs. 1(c) and 1(d) of the manuscript, peaks 10 and 01 would be equally intense; peaks 12, $\bar{1}1$, 21, and $\bar{2}\bar{1}$ would also be equally intense. We do not observe such a highly symmetric diffraction pattern and can, thus, exclude the possibility of a Z=2 skyrmion lattice in our Fe-Gd films. The two fold symmetry of the magnetic diffraction is a clear signature of a lattice of scatterers with out of plane magnetizations that, neglecting polarity, have two fold symmetry.

Our interpretation is that such scatterers are bound pairs of unit topological charge skyrmions of like polarity (based on reasoning explained in the main document). The numerical analysis by Lilliehöök et al[3] lends weight to this, as they demonstrate that it is energetically favorable for skyrmions to form bound pairs rather than Z=2 skyrmions.